**Thermodynamics of hydride formation: Anisotropic size-dependent coupled chemo-thermo-mechanical effects at Ni/NiH interfaces**

**Sourabh Singha and Abhijit Chatterjee***

Department of Chemical Engineering, Indian Institute of Technology Bombay, Mumbai 400076, India

Email: abhijit@che.iitb.ac.in

**Abstract**

Ni nanoparticles are frequently used as catalysts for hydrogenation reactions as well as in hydrogen storage applications. Recently, we have shown that small Ni nanoparticles can absorb hydrogen at < 10 bar pressure to form Ni hydride. During this process, the hydride growth is anisotropic, and a coherent Ni/NiH interface is formed. In order to explain the anisotropy and to comprehensively account for the coupling chemical, mechanical and thermal effects, we develop in this study a simplified chemo-thermo-mechanical enthalpy model for Ni/NiH interfaces in thin films. This model captures the combined influence of extent of hydride formation $x$, temperature $T$, pressure $P$, size effect ($l$), and the α/β interface energy $\lambda$. Two different α/β interface orientations, namely (100) and (111), are investigated. The model is shown to correctly predict the enthalpy and volume changes over a wide range of length scales, from atomically thin layers (~1 nm) to micron scale and larger. This work provides the basis for the development of similar enthalpy models for other solid-state hydrogen storage nanostructured materials where anisotropic growth of hydride phases is also observed.

**TOC image**

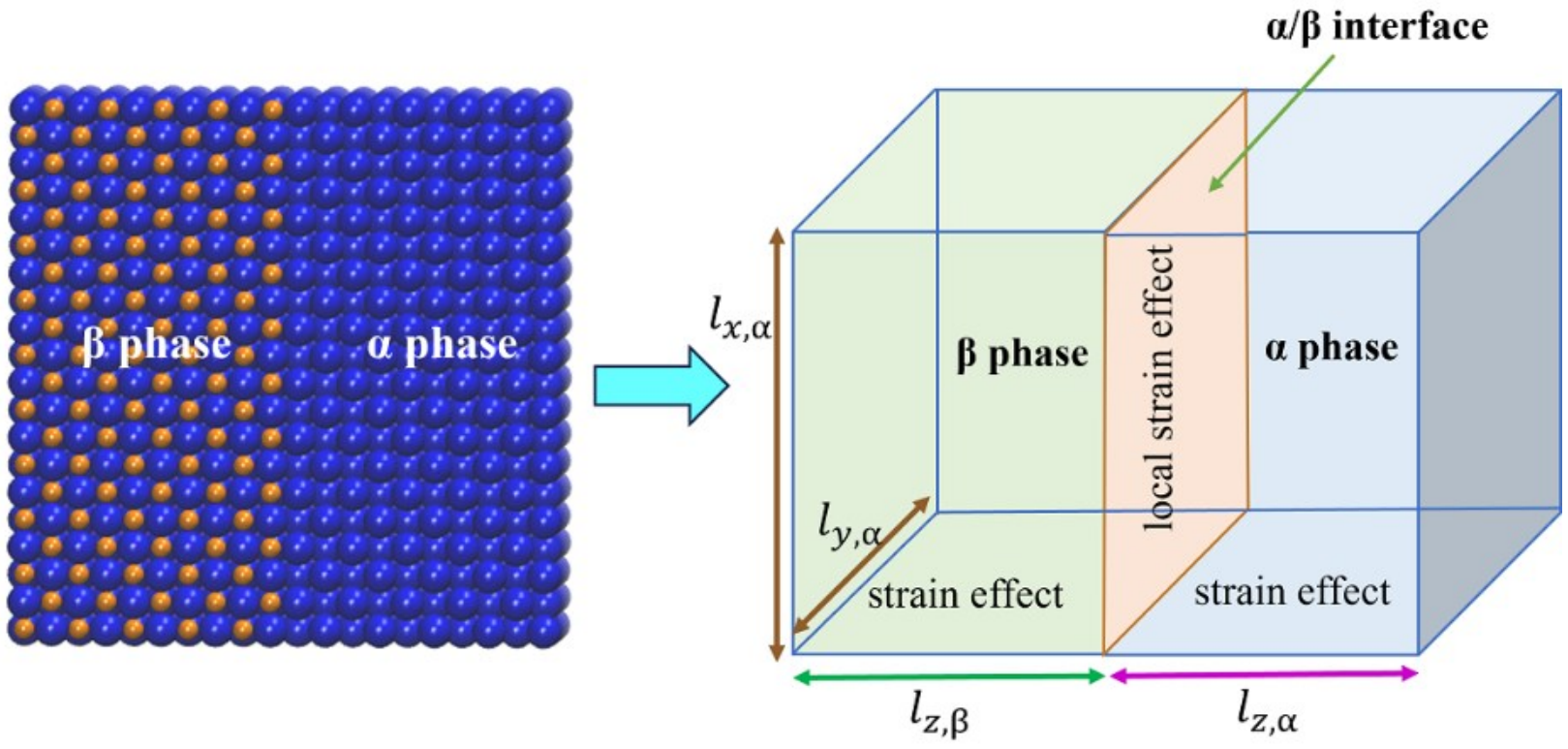

## 1. Introduction

Nickel nanoparticles are frequently used for hydrogenation reactions, offering an alternative to noble metals like palladium or platinum [1–3]. Ni is also used as a catalyst to improve the kinetics of hydrogen storage materials, such as $MgH_2$ [4–10]. Although hydrogen adsorption on Ni surfaces has been extensively studied [11–14], the extent of H absorption in Ni nanoparticles remains largely unexplored. Recently, we have showed that Ni nanoparticles of size < 5 nm are amenable to hydride formation at 10-20 bar hydrogen pressure [15]. This raises important questions about the state of the Ni catalyst during practical operation. Studying the hydrogen absorption and desorption processes in these materials becomes crucial as Ni is often used in the context of hydrogen storage, fuel cells, catalysis, and metal alloy design. Hydrogen incorporation in metals typically causes significant heat release and changes within the host material, such as strain, volume, and even phase transformation [16–20]. In our previous studies, we have investigated the phase behavior of the Ni-H system in bulk and nanoparticle systems, as well as the hydrogen hopping barriers using atomic scale simulations [15,21,22]. Here we comprehensively examine the heat release and volume expansion during hydrogen absorption [23].

Based on the known hydrogen hopping rates in Ni [21], it can be concluded that hydride growth is characterized by a slow moving α/β interface [24]. The α-phase offers low H solubility, whereas, the hydride or the β-phase is rich in H [25][26][27]. Depending on the particle size, it may take minutes to several hours for the hydriding process to complete. Fig. 1a shows a schematic. This allows us to treat the hydriding process as a quasi-equilibrium process. Complete hydride formation in nickel results in a 19% volume expansion [21]. In order for a coherent lattice to form, the α/β region becomes elastically strained. To calculate the heat released as a function of the β phase formed, chemo-thermo-mechanical effects need to be accounted for.

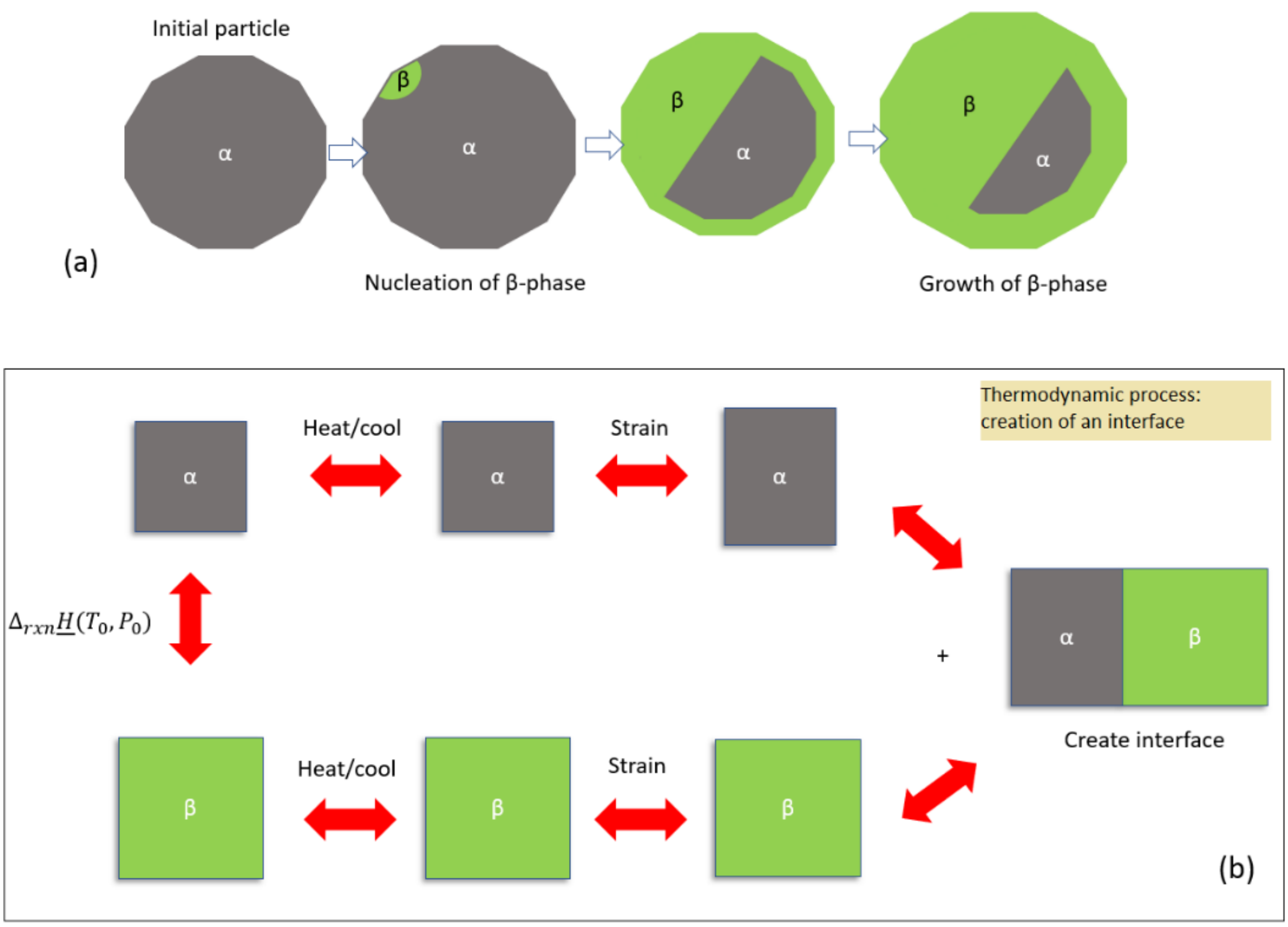

**Fig. 1.** Schematic showing (a) hydrogen incorporation (α→β transition) in a nickel particle, (b) thermodynamic path for creation of an α/β interface in a nickel hydride system.

Coupled chemo-thermo-mechanical effects have also been highlighted by others. Traisnel et al. highlighted the significance of strain during hydrogen incorporation in Ni, and how volume expansion enhances hydrogen solubility [28]. In other metal hydrides systems as well, such as, PdH, $LaNi_5$ [29], $MgH_2$ [30], $LiBH_4$ [31], the thermodynamic properties of the α/β system are a function of the chemical, mechanical and thermal effects. Given this context, it is important to build material-specific thermodynamic models that incorporate coupled chemo-thermo-mechanical effects.

Consider the example of the continuum elasticity theory, which can be used to predict the macroscopic elastic deformation behavior of a layered α/β system [32]. Continuum models typically employ a set of elastic constants. These elastic constants are constant parameters, and ignore the effect of temperature, interfaces, and dimensions of the materials. Atomistic scale models can be used to assess the coupled effects, and then build appropriate continuum models. Similarly, the growth of a new phase involves the creation of interfaces. Unlike liquids, different crystallographic planes are present in the growing crystalline phases. Planes associated with low interfacial energy are preferred, and the relative energies usually dictate the growth direction. The material feature size affects the thermodynamics of interfaces, particularly in nanostructured materials, where the high surface to volume ratio influences the interface energy [33]. In such contexts, atomistic simulations can help decouple chemo, mechanical and thermal effects.

We develop a continuum enthalpy model for the α/β interface of the $NiH_x$ system, where $x$ is the extent of hydride formation, $0 \leq x \leq 1$. The study also examines the size effect in a layered nickel hydride system. The computational protocol to bridge the gap between atomistic simulation and continuum models is useful for other solid-state hydrogen storage materials as well.

The outline of this paper is as follows. In Section 2, we explain the theory and section 3 presents computational methods. Section 4 mentions the results and discussion. Finally, Section 5 describes our conclusions and findings. The symbols used are listed in Table 1.

**Table 1. List of symbols used.**

| Nomenclature | | |
|---|---|---|
| Symbol | Description | Units |
| α | Ni | |
| β | NiH | |
| $C_{P,i}$ | Heat capacity for phase $i$, $i$ is $\alpha$ or $\beta$ | eV/(Ni atom.K) |
| $\Theta_i$ | Thermal expansion coefficient for phase $i$, $i$ is $\alpha$ or $\beta$ | 1/K |
| $\underline{H_i}$ | Change of molar enthalpy for phase $i$, $i$ is $\alpha$ or $\beta$ | eV/(Ni atom) |
| $\Delta\underline{H}_{heat,i}$ | Thermal energy for phase $i$, $i$ is $\alpha$ or $\beta$ | eV/(Ni atom) |
| $\underline{\Delta H}_{el,i}$ | Elastic energy for phase $i$, $i$ is $\alpha$ or $\beta$ | eV/(Ni atom) |
| $\lambda$ | Interfacial energy | eV/(Å$^2$) |
| $A_{int}$ | (Molar) interfacial area* | Å$^2$/(Ni atom) |
| $\gamma$ | Surface energy | eV/(Å$^2$) |
| $A_{surf}$ | (Molar) surface area* | Å$^2$/(Ni atom) |

*Molar area refers to the area per mole of Ni.

## 2. Theory

Fig. 1a shows the different stages of hydrogen absorption in a Ni particle. Initially, the hydride nucleates at the particle surface. The α/β interface moves, as the hydride phase (green color) grows. We simplify the picture by focusing on the interfacial region in a 2D film (Fig. 1b). Fig. 1b shows a thermodynamic path for Ni/NiH interface formation at a temperature $T$ and pressure $P$. To create a coherent interface, we first bring the α-phase to the desired temperature. Next, we strain the material at a constant temperature. Similar steps are performed for the bulk β-phase. The final step is the formation the α/β interface. The interface has an associated energy, $\lambda$. The relative thickness of the α- and the β-regions is related to the extent of hydride formation, $x$. We determine the molar enthalpy function, $\underline{H}(x, T, P, l, \lambda)$, for such an α/β interfacial system. The final dimensions $\{l\}$ of the α- and the β-regions, are dictated by minimization of the free energy.

### 2.1. Chemo-thermo-mechanical model

In experiments, the α-phase can contain some H (<2%), whereas the β-phase can have a H:Ni ratio of 0.8-1:1. We assume that the α-phase is pure Ni, and the β-phase is pure NiH, i.e., H:Ni=1:1 in the β-phase, as shown in Fig. 2(a, b). By making this approximation, we can ignore the hydrogen configurational aspects. Hydrogen incorporation in Ni proceeds along the z-direction.

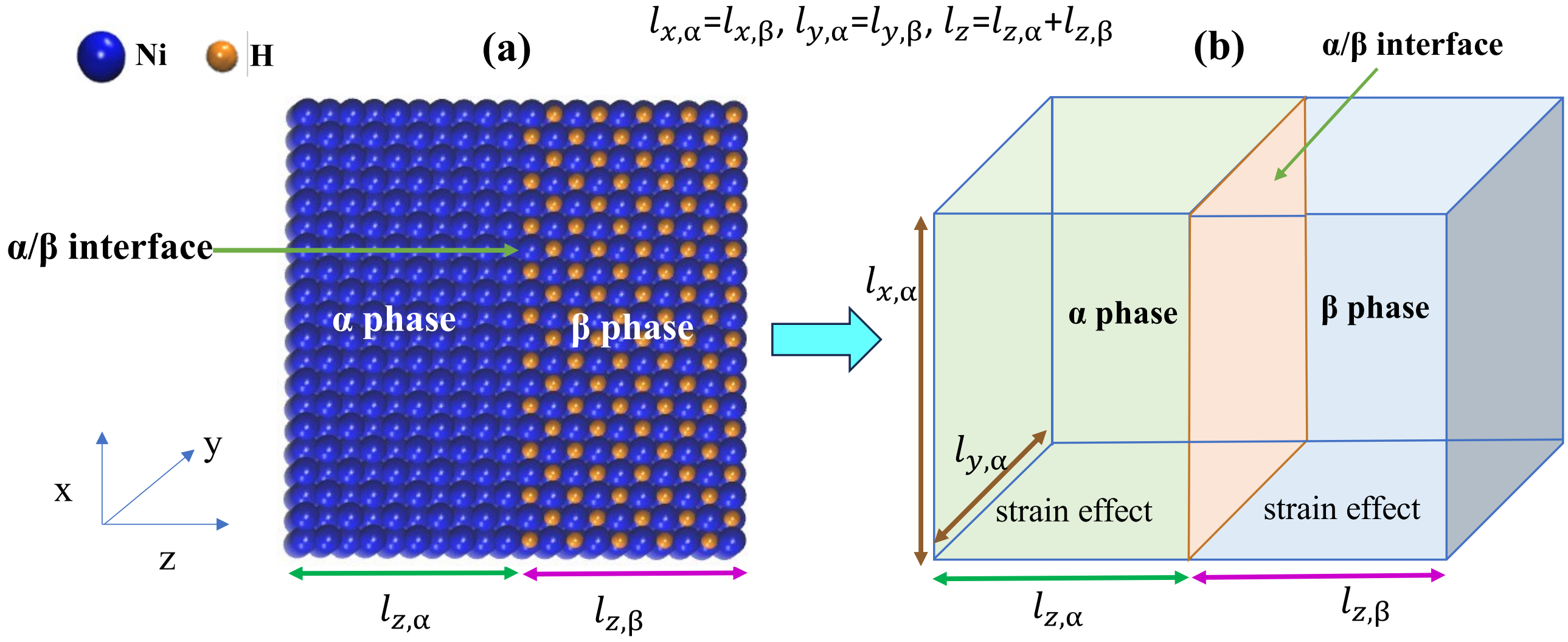

**Fig. 2.** (a) Schematic shows the atomistic hydride system, where the α, β phase and α/β interface are mentioned. Here, the α phase is pure Ni and the β phase is pure NiH. (b) Continuum system considered for the elastic energy calculation. The length parameters $l_{z,\alpha}$, $l_{z,\beta}$, $l_{x,\alpha}$ and $l_{y,\beta}$ are shown more clearly in panel b.

The molar enthalpy of the metal/metal hydride system can be written as

$$\underline{H}(T,P,x,\{l\},\lambda) = x_\alpha \underline{H_\alpha}(T,P,\{l\}_\alpha) + x_\beta \underline{H_\beta}\big(T,P,\{l\}_\beta\big) + A_{int}\lambda + A_{surf,\alpha}\gamma_\alpha + A_{surf,\beta}\gamma_\beta \quad (1)$$

where, $\underline{H_\alpha}$and $\underline{H_\beta}$ are the contributions from the bulk phase, $x_\alpha$ and $x_\beta$ denote the fraction of each phase, $\lambda$ is an interfacial energy, and $\gamma$ is surface energy. $A_{int}$ and $A_{surf}$ are interface and surface area, respectively, normalized by the number of Ni atoms so that the units are consistent. The interface and surface area are reported in ($Å^2/Ni\ atom$), whereas $\lambda$ and $\gamma$ have units of $eV/Å^2$.

The bulk phase contributions for phase $i$ ($i = \alpha,\beta$), i.e., the first and second terms in the right-hand side of equation (1), are calculated as

$$\underline{H_i}(T,P,\{l\}_i) = \underline{H_i}(T_0,P_0) + \Delta\underline{H}_{heat,i} + \Delta\underline{H}_{el,i} \quad (2)$$

using the thermodynamic path of Fig. 1(b). No surfaces/interfaces are present. Here $\underline{H_i}(T_0,P_0)$ is the molar enthalpy of the starting system in Fig. 1(b). $\underline{H_i}$ has units of $\frac{\text{eV}}{\text{Ni atom}}$. $T_0$ is 300 K and $P_0$ is 1 bar. Pressure effects are negligible in solids. Let $N_\alpha$ and $N_\beta$ be the number of Ni atoms in the α and β phase, respectively, such that $N$, which is given by $N = N_\alpha + N_\beta$, is fixed. The total enthalpy of the system, i.e., $N\underline{H}$, is measured with the help of MD simulations.

## 3. Computational methodology

### 3.1. System preparation

The fcc unit Ni system has lattice parameters $a = b = c =$ 3.52 Å and an angle $\alpha = \beta = \gamma = 90°$ with space group Fm$\bar{3}$m (No.225) [34]. Hydrogen occupies the octahedral interstitial site [28,35]. The bulk nickel hydride ($NiH_x$, x=0-1) possesses an NaCl structure with a lattice constant of 3.738 Å (19% vol. expansion) [25,36]. For single phase calculations involving Ni or NiH, the system contains a 10×10×10 supercell and periodic boundary conditions (PBC) are imposed along the x, y, and z direction.

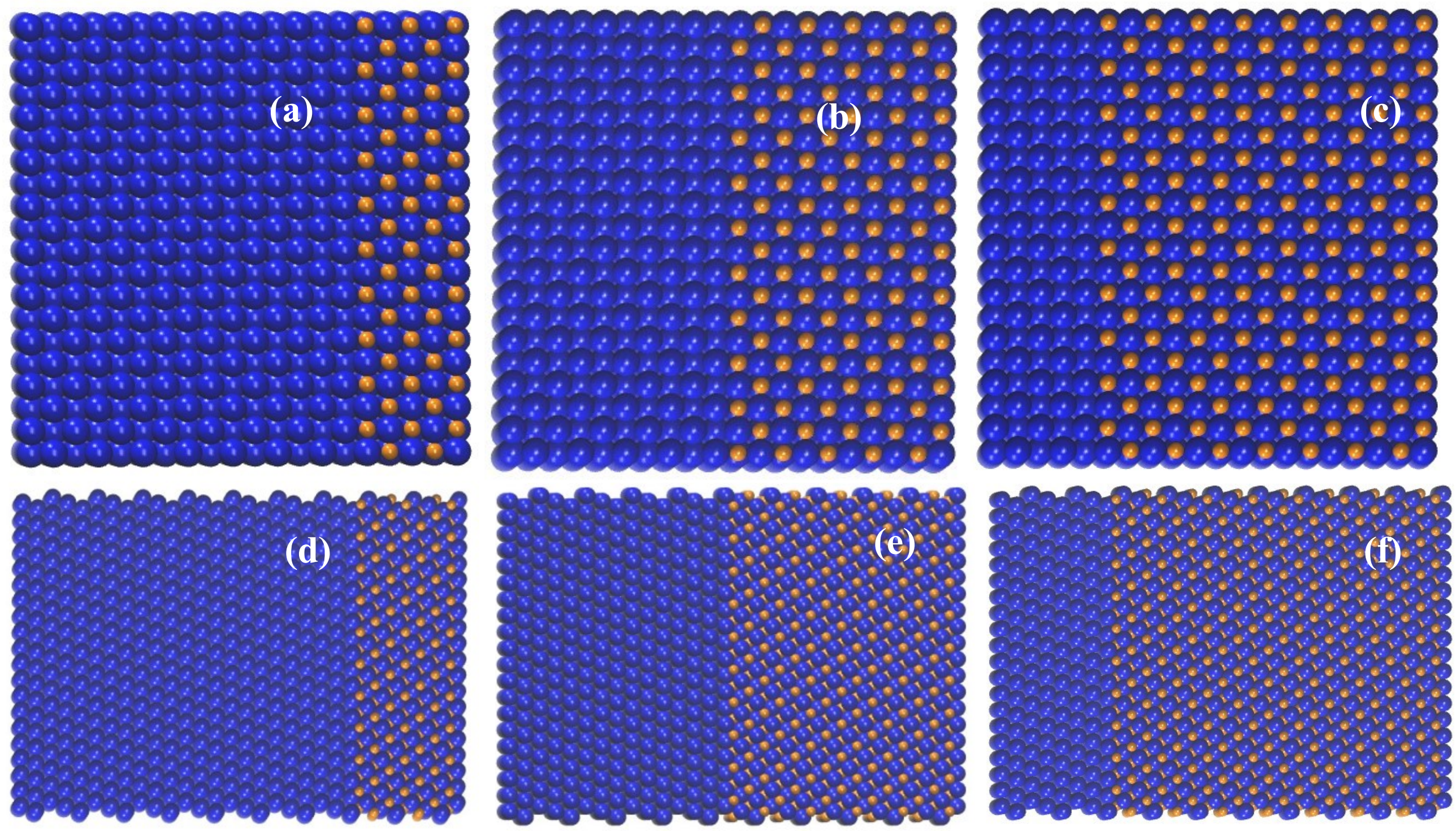

**Fig. 3.** The α/β interface in bulk $NiH_x$ system. Top, and bottom rows are the {100}, and {111} interface, respectively. Left, middle and right columns have x being 0.25, 0.50 and 0.75, respectively. Here Ni=blue, H=orange.

For Ni/NiH systems (α/β interface), we have used supercells of different sizes, such as 4×4×2, 4×4×4, 4×4×40, 4×4×100, 4×4×180 and 4×4×260 repeat units to probe the size effect. Two interfaces, namely, in the {100}, and {111} directions, as shown in Fig. 3, are studied to calculate interfacial energy. MD simulations are performed with x=0, 0.25, 0.5, 0.75 and 1 in the overall $NiH_x$ with these interfacial systems to mimic the movement of hydrogen in the z-direction.

It is known that cracks and powdering are observed in large particles during hydriding. This happens due to nucleation of defects, such as dislocations. The probability of dislocation formation decreases dramatically for smaller system sizes, such as in nanoparticles[37]. Narayan et al. [38] showed the presence of a coherent interface in Pd nanoparticles as they form hydride by using scanning transmission electron microscopy (STEM). For this reason, we believe that the coherent lattice approximation used in this study should be generally valid for nanostructured materials. Additionally, it should be point out that in our simulations the dimensions of the Ni and NiH regions are fairly small and the molecular dynamics simulations are short. Therefore, dislocation and crack formation are unlikely in such simulations.

### 3.2. Molecular dynamics

Molecular dynamics (MD) can accurately capture thermal expansion, elastic effects, interface energies. MD simulations are performed using the open source code Large-scale Atomic/Molecular Massively Parallel Simulator (LAMMPS) package [39]. The periodic boundary conditions (PBC) are imposed in all directions. NiAlH embedded atom method (EAM) interatomic potential, developed by Baskes et al. [34,40], is used. This EAM potential yields good agreement with experimental measurements for the Ni-NiH system, as shown previously [15,21,22,40–42]. In our previous studies, we have investigated the phase behavior of the Ni-H system in bulk and nanoparticle systems, as well as the hydrogen hopping barriers using atomic scale simulations [15,21,22]. Of particular importance is the heat release and volume expansion during hydrogen absorption [23], which are discussed in this study.

MD simulations are performed in the NPT or NVT ensemble depending on the requirement. A timestep of 0.5 fs, pressure of 1 bar, and a temperature of 300-800 K is used. The Nosé-Hoover thermostat [43] and barostat [44] are used to maintain the set-point temperature and pressure, respectively. For NPT-MD simulations, the simulation cell shape was allowed to change anisotropically, i.e., the dimensions in the x-, y- and z-directions are allowed to change independently.

The initial 50 ps of the MD trajectory was discarded to ensure thermal and mechanical equilibration. Thereafter, the MD trajectory was extended for 700 ps. Block averaged behavior in Supplementary Material confirms proper equilibration has happened. The trajectory was employed to compute average energy, which were later utilized for the calculation of compute heat capacity and elastic constants.

### 3.3. Energy minimization

Energy minimization is used to prepare relaxed structures at 0 K. These calculations were again performed using the LAMMPS package. Periodic boundary conditions (PBC) are imposed in all directions. The energy minimization is performed using conjugate gradient algorithm (CGA) with energy tolerance of $10^{-10}$ eV and force tolerance of $10^{-9}$ eV/Å. Bulk properties such as lattice parameter, cohesive energy, elastic constant and heat of formation at 0 K were obtained.

## 4. Results and discussion

The enthalpy model (equation (1)) contains various contributions. This section we evaluate the related parameters step-by-step as follows: (i) bulk properties at 0 K, (ii) temperature effects, (iii) development of a temperature-dependent elastic model, (iv) interfacial energy as a function of strain and temperature, and finally, (v) development of the full enthalpy model.

### 4.1. Lattice parameter, cohesive energy, elastic constants, and heat of formation at 0 K

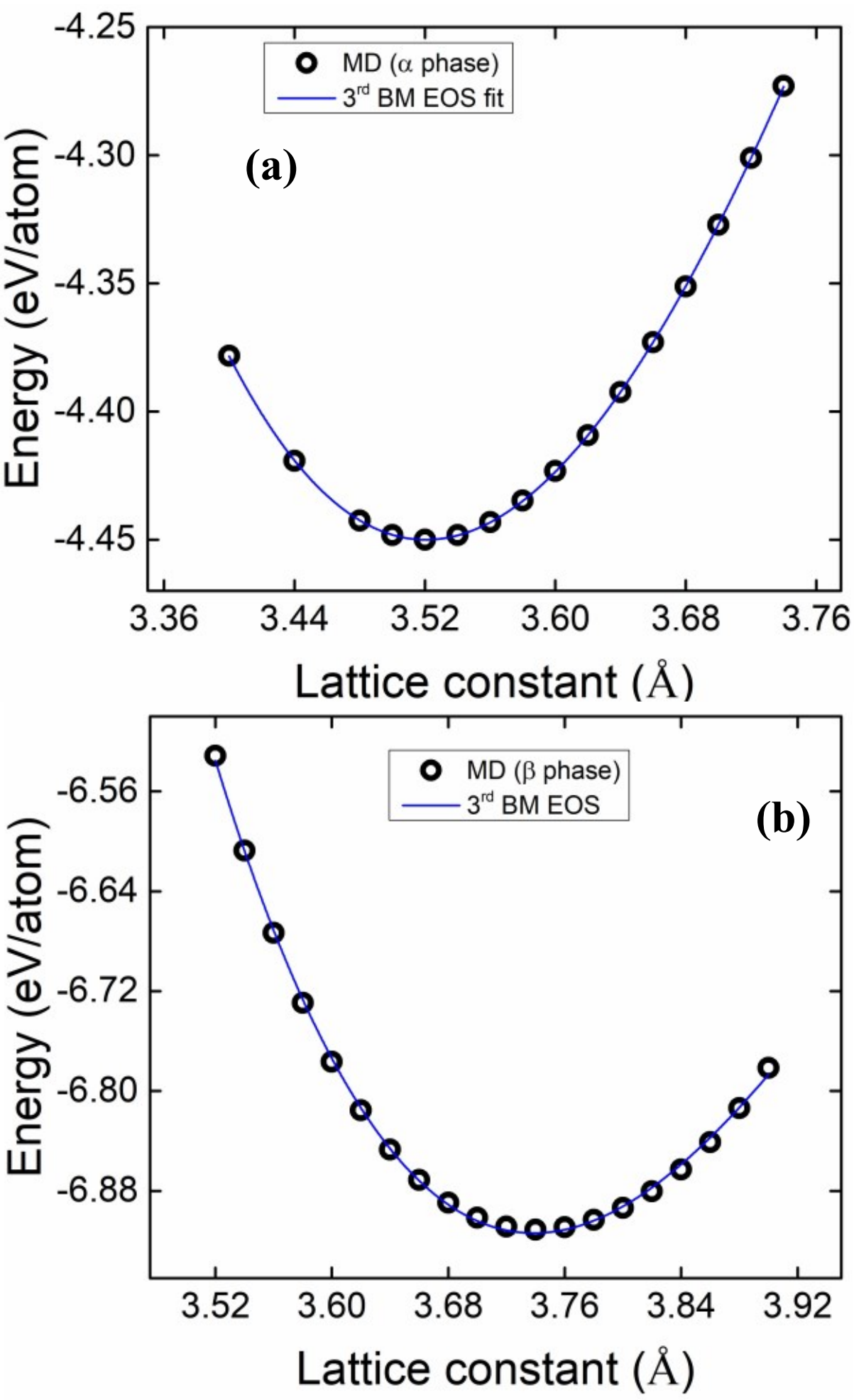

**Fig. 4.** Energy of (a) Ni (α phase) and (b) NiH (β phase) as a function of lattice constant at 0 K. The open circle symbols are energy minimization calculation, while the curve denotes a fit to the Birch-Murnaghan (BM) equation of state (EOS). Here, energy is reported in units of eV per Ni atom.

Lattice parameter, bulk modulus of the bulk Ni and bulk NiH are calculated. The equilibrated lattice parameters of the α and β phases were obtained by cell relaxation and energy minimization and agree well with the literature values (see Table 2). Fig. 4(a) and (b) yield the cohesive energy at 0 K for the Ni and NiH phase. The computed cohesive energy for Ni is -4.45 eV and its lattice parameter is 3.52 Å, as shown in Fig. 4a. This compares well with the known experimental value of -4.44 eV [45,46] and previously reported simulation results for EAM is -4.45 eV [40], first-principle -4.46 eV [47], and MEAM -4.45 eV [48]. The cohesive energy for NiH is -6.91 eV is mentioned in Fig. 4b, and the lattice constant is 3.74 Å.

**Table 2: Bulk properties of α and β phase. Literature value is mentioned in the bracket.**

| Phase | Lattice parameters (Å) | Cohesive energy (eV) | Bulk modulus (eV/$Å^3$) |
|---|---|---|---|
| Ni (α) | a=b=c= 3.52 (3.52) [34] | -4.45 (-4.44 [45,46]) | 1.141 (1.131 [49,50]) |
| NiH (β) | a=b=c= 3.74 (3.735 [25]) | -6.91 | 1.314 (1.236,1.327 [51–53]) |

Fig. 4 also provides the bulk modulus of Ni and NiH bulk system. The bulk modulus was calculated by fitting the Birch-Murnaghan (BM) equation of state (EOS) [54]. The computed bulk modulus of Ni is 1.141 eV/$Å^3$ which is close to the literature value of 1.131 eV/$Å^3$ [49,50] (see Table 2). The computed bulk modulus of NiH is 1.314 eV/$Å^3$. Experimental values for NiH lie between 1.236 to 1.327 eV/$Å^3$ [51–53].

The heat of formation for bulk NiH is calculated as $\Delta_f \underline{H} = \frac{H_{NiH} - H_{Ni} - \frac{n_H}{2} H_{H_2}}{n_{Ni}}$ is found to be $-2.46\ \mathrm{eV/Ni\ atom}$, which matches with values reported in literature [22]. Here, $n_H$ is the number of hydrogen atom, $H_{NiH}$ is the energy of pure Ni-H system, $H_{Ni}$ is the total energy of pure Ni system, and $H_{H_2}$ is energy of hydrogen molecule.

### 4.2. Temperature effect: heat capacity and thermal expansion coefficient

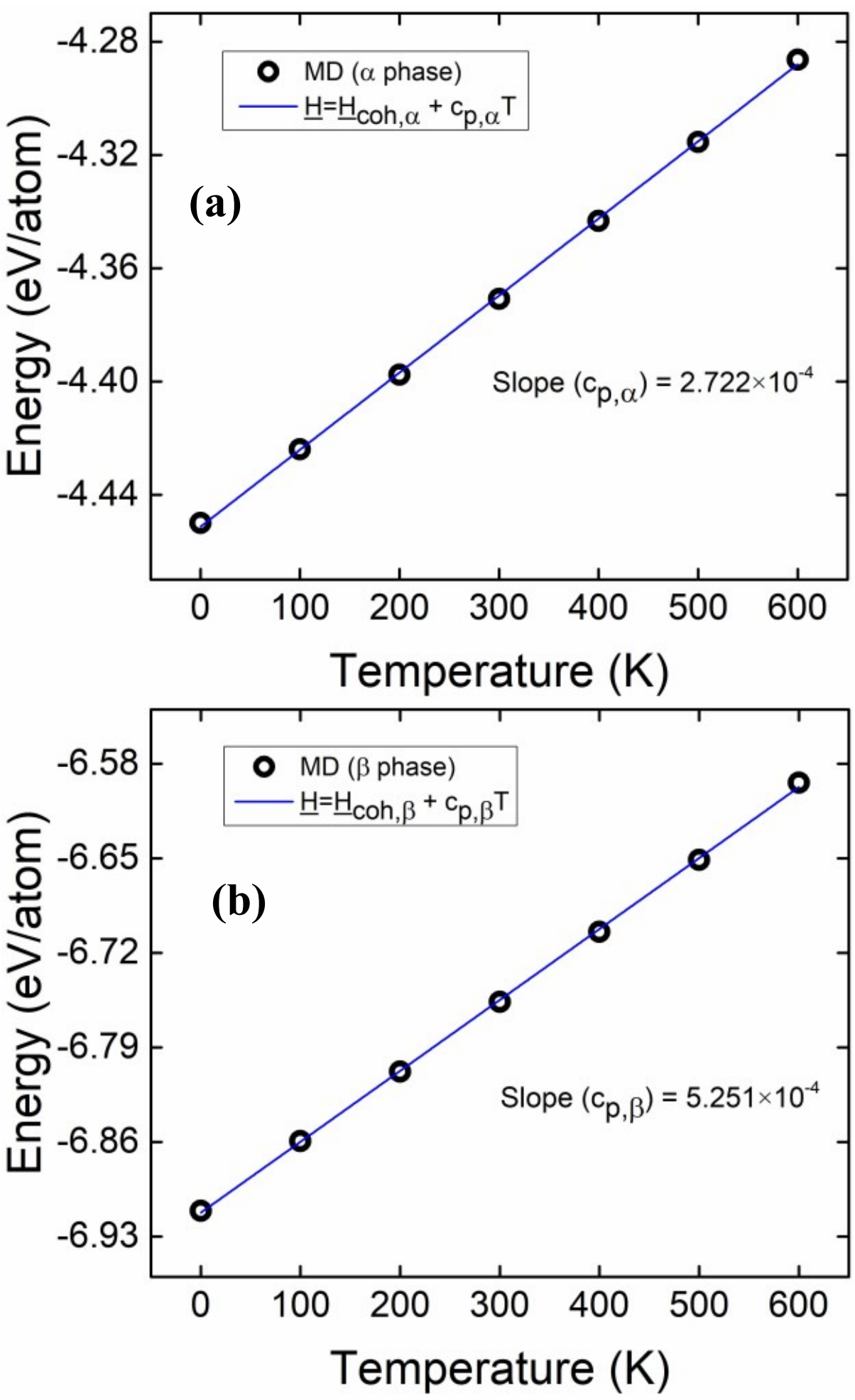

**Fig. 5.** Plot shows total internal energy as a function of temperature for (a) α phase and (b) β phase for nickel hydride system. Here, energy is reported in units of eV per Ni atom.

MD simulation of Ni and NiH were performed in the NPT ensemble with a 10×10×10 supercell at temperatures $T$ = 100-600 K. Upon heating, the molar enthalpy of phase $i$ follows

$$\underline{H_i}(T, 1\ bar) = \underline{H_i}(T_0, 1\ bar)\ +\ C_{P,i}\ (T - T_0). \quad (3)$$

Here, $C_{P,i}$ is the heat capacity of a single phase $i$, where $i$=α, β. The first 50 ps of the simulation is discarded to ensure thermalization. Subsequent trajectory provides the time-averaged energy and lattice parameter, which are plotted in open circles in Fig. 5 and 6, respectively. The temperature-dependence observed in the calculated energy is consistent with equation (3). Since the $PV$ term can be ignored due to its small value, the enthalpy is approximately equal to the internal energy. The computed heat capacity of Ni and NiH ($C_{P,\alpha}$ and $C_{P,\beta}$) is $0.2722$ and $0.5251\ \frac{\text{meV}}{\text{K Ni atom}}$, respectively. See Table 3 for comparisons to experiments. Our simulation results are in good agreement with the experimental heat capacity [55][56].

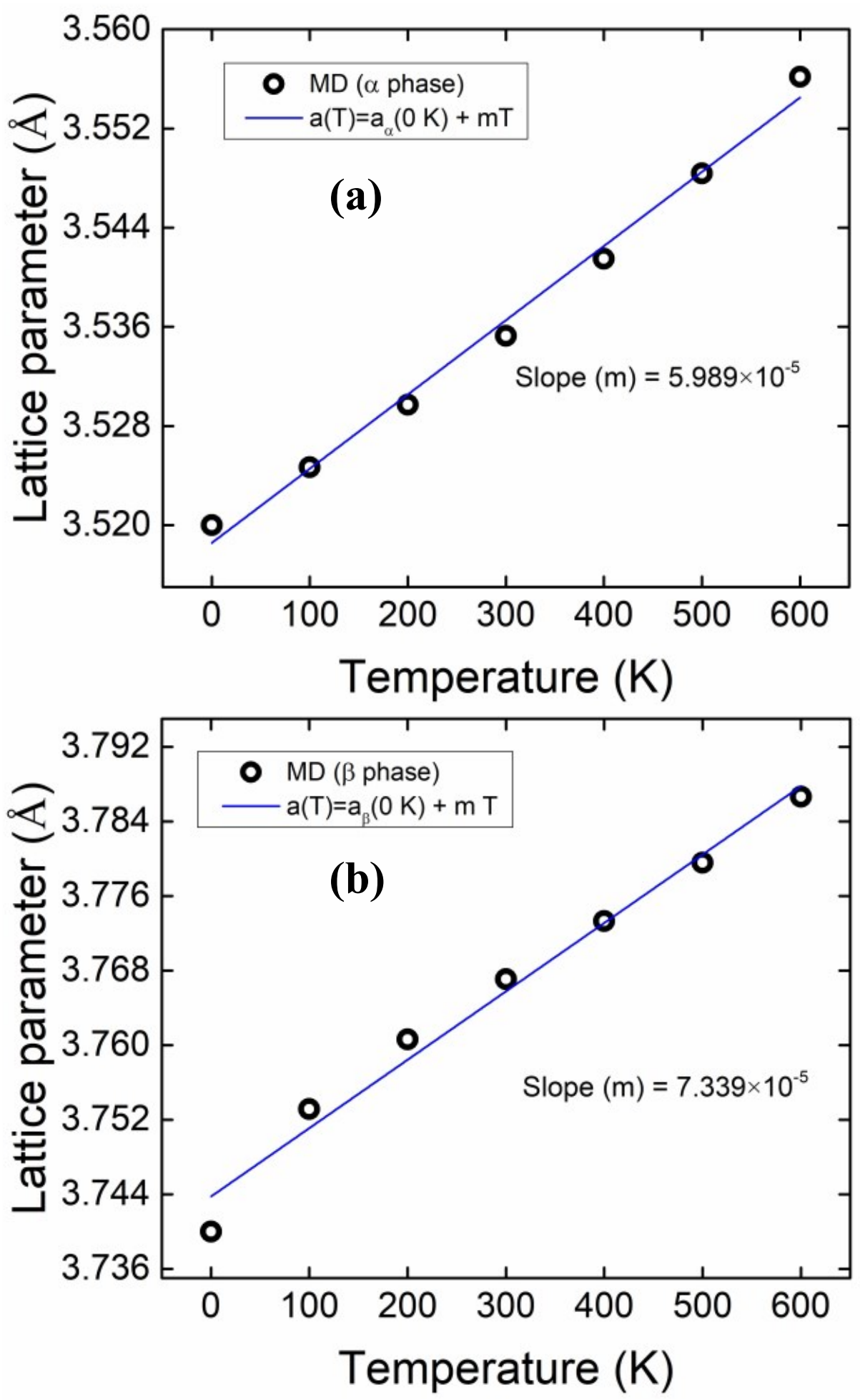

**Fig. 6.** Temperature-dependent lattice parameters of (a) bulk Ni and (b) bulk NiH phases. Open circles denote MD data. Blue line denotes equation (4).

Next, we consider thermal expansion of a single-phase material. Let $a_i$ denote the lattice parameter at temperature T, for i=Ni or NiH. Assuming the thermal expansion coefficient to be a constant,

$$a_i(T, 1\ bar) = a_i(T_0, 1\ bar) + m_i(T - T_0). \quad (4)$$

The volumetric thermal expansion coefficient (ϴ) is $\Theta_i = 3m_i$ where $i$=α, β. The thermal expansion coefficient of Ni and NiH phase are also obtained from MD simulation in the NPT ensemble (see Table 3). Fig. 6 shows the equilibrium lattice parameters for the α and β phases vs. temperature. The blue line represents equation (4) obtained via linear regression. Correlation coefficient $R^2$ is found be 0.99 and 0.98 for Ni and NiH respectively can be seen in Fig. 6 (a, b). Small variations in thermal expansion coefficient with respect to temperature are ignored for simplicity, and later the interfacial properties are shown to be accurately estimated with this approximation.

**Table 3: Heat capacity and linear thermal expansion coefficient of α and β phase is obtained from Fig. 5 and 6. Literature value is mentioned in the bracket.**

| Thermal property | α phase | β phase |
|---|---|---|
| $C_p$ (meV/K. Ni atom) | 0.2722 (0.27 [55]) | 0.5251 |
| ϴ (1/K) | $5.989 \times 10^{-5}$ ($5.6 \times 10^{-5}$ [57]) | $7.339 \times 10^{-5}$ |

**4.3. Elastic model and strain energy contribution to the enthalpy**

Next, we focus on a situation where a single phase, either α or β, is elastically strained. The stress tensor $\sigma$ in a material phase can be written in terms of the strain tensor $\epsilon$ as

$$\sigma_{kl} = \sum_{m=1}^{3} \sum_{n=1}^{3} C'_{klmn} \epsilon_{mn} \quad (5)$$

where, $k, l = 1,2,3$ and $C'_{klmn}$ is the fourth order elastic rotated stiffness tensor[58]. The rotated stiffness tensor is obtained in the index notation as $C'_{klmn} = Q_{kp}Q_{lq}Q_{mr}Q_{ns}C_{pqrs}$, where $Q_{kp}$ denotes the directional cosine[58]. The elastic enthalpy ($H_{el}$, in eV) for the bulk phase is [58]

$$\Delta H_{el} = \frac{1}{2}\sigma_{kl}\epsilon_{kl}V. \tag{6}$$

Here, $V$ is the unstrained volume. While the stress tensor components can be related to the strain tensor using the fourth order elastic stiffness tensor (equation (5)), it is more convenient to use the Voigt notation. In a fcc and NaCl crystal system, there are three independent elastic constants $C_{11}$, $C_{12}$, and $C_{44}$ [59]. Here the subscripts are written according to the Voigt notation. For the system shown in Fig. 2, the shear strain is zero. Note that Ni and NiH have different elastic constants. The elastic energy per Ni atom at α and β phase appearing in equation (2) is computed as

$$\Delta \underline{H}_{el,\alpha} = \frac{\Delta H_{el,\alpha}}{N_\alpha} \tag{7}$$

and

$$\Delta \underline{H}_{el,\beta} = \frac{\Delta H_{el,\beta}}{N_\beta}. \tag{8}$$

Temperature appears in the definition of strain as follows. Ni and NiH can undergo thermal expansion. Consider the α phase. Let us write the length along x, y, z direction as $l_{x,\alpha 0}\,(T, P_0)$, $l_{y,\alpha 0}(T, P_0)$, and $l_{z,\alpha 0}(T, P_0)$. The subscript 0 implies zero mechanical strain. The $V$ appearing in equation (6) is $V_{\alpha 0}$=$l_{x,\alpha 0}l_{y,\alpha 0}l_{z,\alpha 0}$. The mechanical strain along $\xi = x, y, z$ in equation (6)-(7) is calculated as

$$\epsilon_{\xi\xi,\alpha} = \frac{l_{\xi,\alpha}}{l_{\xi,\alpha 0}} - 1. \tag{9}$$

Similar expressions are valid for the NiH phase. Shear strain terms are absent in our simulation setup. Elastic constants determined at 0 K are listed in Table 4. Elastic constant of Ni is in good agreement with experiment and previous simulation studies as shown in Table 4. The elastic constant of NiH is not available in literature, however the tendency is that $C_{11}$ and $C_{44}$ values are decreasing and $C_{12}$ values are increasing as x increases is reported in literature [60] and our simulation show similar trends. Elastic property at different temperature is calculated for pure Ni and pure NiH system is shown in supplementary Fig. S4.

**Table 4: Computed elastic constants at 0 K. Experimentally and other simulation measured values are shown in parenthesis. Literature value is mentioned in the bracket.**

| Elastic constant | Ni | NiH |
|---|---|---|
| $C_{11}$ (eV/Å$^3$) | 1.539 (1.53 [59][60]) | 1.488 |
| $C_{12}$ (eV/Å$^3$) | 0.9272 (0.928 [59][60]) | 1.143 |
| $C_{44}$ (eV/Å$^3$) | 0.779 (0.778 [59][60]) | 0 |

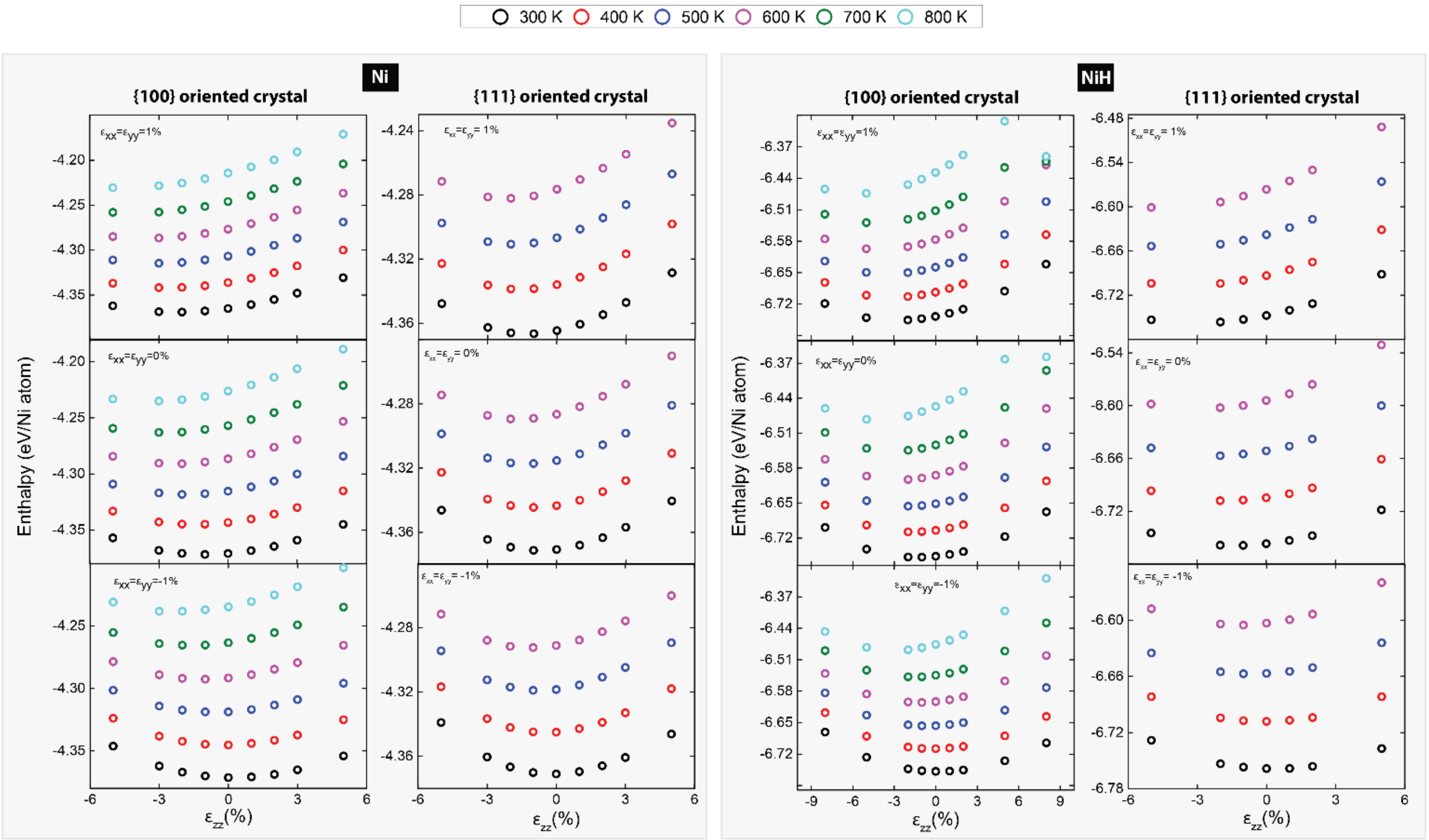

**Fig. 7.** Molar enthalpy for bulk Ni and NiH at different elastic strain and temperature with {100} and {111} plane. Here for {100} plane temperature varies from 300 K to 800 K and for {111} plane 300 K to 600 K. Initially more data was generated for {100} plane for the prediction of enthalpy model so temperature range was taken 300 K to 800 K. The strain is varied in nickel for {100} plane in x, y direction to -5 to 5 % is reported in Supplementary Fig. S6. The strain is varied in nickel hydride for {100} plane in x, y direction to -8 to 2 % is reported in Supplementary Fig. S7.

To quantify the coupled strain and temperature effect, we perform MD simulations in the NVT ensemble. The crystal is oriented in the {100} direction. Fig. 7 shows the results for the Ni and NiH phases, respectively for {100} and {111} plane. Each panel considers a particular value of $\epsilon_{xx} = \epsilon_{yy}$. Each open circle denotes a result from an MD calculation, wherein additionally the temperature and $\epsilon_{zz}$ are specified. Thus, $\epsilon_{xx} = \epsilon_{yy} = \epsilon_{zz} = 0$ refers to the volume relaxed, mechanically unstrained system. The reader will realize that the enthalpy function $\underline{H}$ is not minimized at this condition, only the Gibbs free energy $\underline{G}$ is (not shown). For simplicity, the entropic terms are ignored here. This amounts to writing $\underline{G} \approx \underline{H}$. Entropy terms were shown to be small in our earlier work [22]. In Fig. 7, $\epsilon_{ii}$, $i = x, y, z$ is varied between -5% to 5% in anticipation of the strain values encountered later in the α/β systems.

Since the elastic constants are positive-valued, the constant temperature curves in each panel have a positive curvature. As the temperature is increased, the curves shift vertically up due to the heat capacity term in equation (2). However, the key point of the figure is the change in the curvature with temperature, which suggests a coupling between temperature and strain effects. Similar behavior is observed with NiH (see Fig. 7). This coupling is captured by writing the heat capacity and elastic constant parameters with temperature and strain corrections, such that

$$C_{P,i} = C_{P,0} + C_{P,T}\, T + C_{P,\epsilon}(\epsilon_{xx} + \epsilon_{yy} + \epsilon_{zz}) \quad (10)$$

$$C_{11,i} = C_{11,0} + C_{11,T}T + C_{11,\epsilon}(\epsilon_{xx} + \epsilon_{yy} + \epsilon_{zz}) + \ C_{11,T\epsilon}(\epsilon_{xx} + \epsilon_{yy} + \epsilon_{zz})T$$

$$C_{12,i} = C_{12,0} + C_{12,T}T + C_{12,\epsilon}(\epsilon_{xx} + \epsilon_{yy} + \epsilon_{zz}) + C_{11,T\epsilon}(\epsilon_{xx} + \epsilon_{yy} + \epsilon_{zz})T.$$

The first term on the right-hand side of the equation (10) is the regarded as the value at zero temperature and strain, whereas the remaining terms capture the coupling. All parameters in equation (10) are determined via regression. This is done separately for the Ni and NiH.

The simplest model is one where the corrections are set to zero (no coupling present). The heat capacity determined from Fig. 5, and the elastic constants at 0 K. Fig. 8(a) shows a parity plot obtained using such a model (see red open circles). The enthalpy from MD calculations in NVT ensemble are mentioned along the horizontal axes of Fig. 8, and the vertical axis corresponds to equation (2) for the same conditions. The agreement between such an enthalpy model and MD results is poor. Adding the corrections (see black plus symbols) improves the model accuracy (see Fig. 8(a) and (b)).

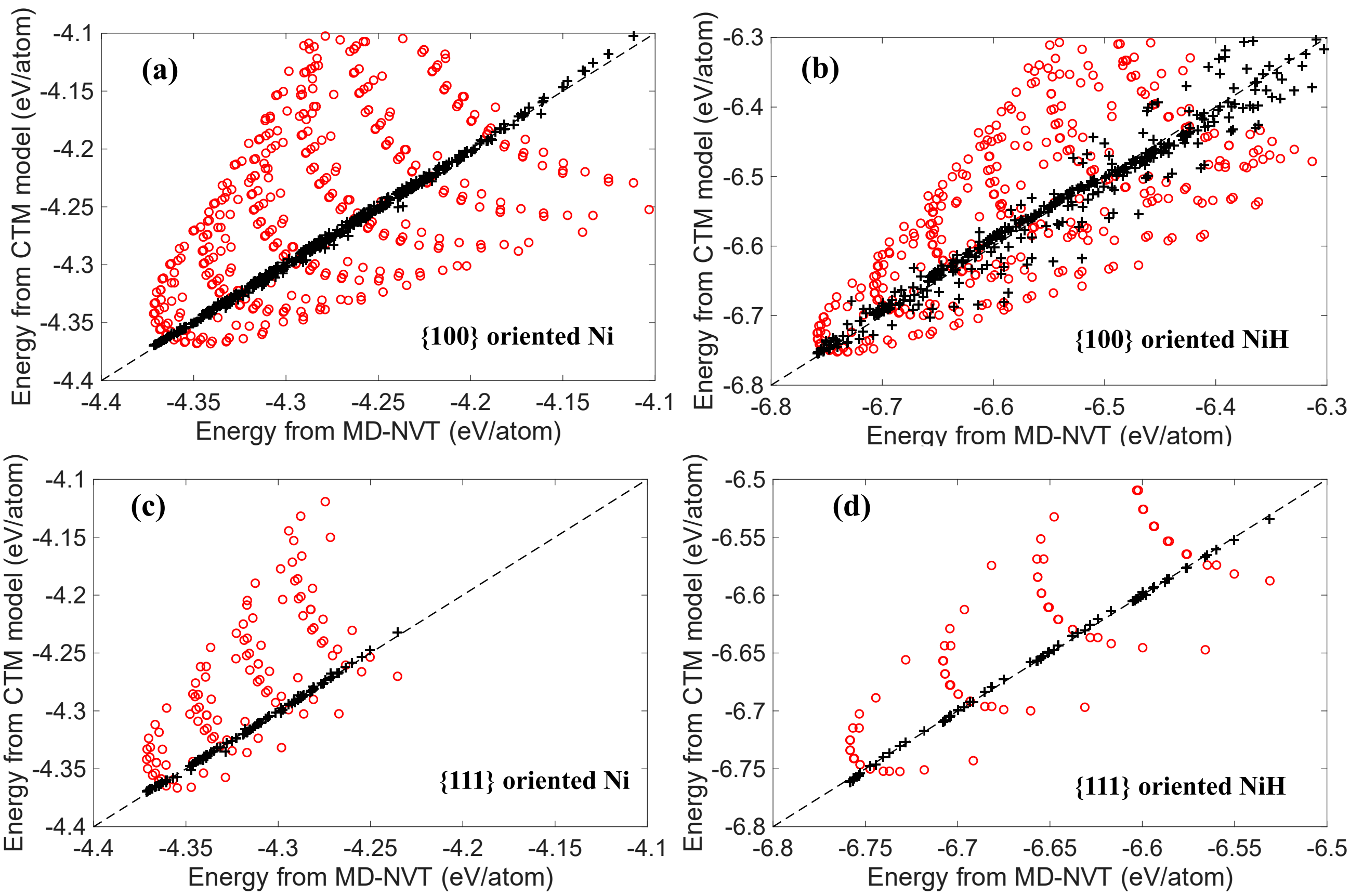

**Fig. 8.** Chemo-thermo-mechanical (CTM) enthalpy model showing the importance of coupled temperature and strain effects in (a) α and (b) β phase when the bulk crystal is {100} oriented. Corresponding results for {111} orientation: (c) α phase, (d) β phase. Coupling parameters were obtained from panels (a) and (b) using equation (10). These parameters along with tensor transformation are directly used in panels (c)-(d). Red circles: No coupled temperature and strain corrections included. Black plus symbol: Corrections included. Here, energy is reported in units of eV per Ni atom.

**Table 5. Elastic constant, heat capacity and their corrections (see equation (10)) for phase $i$, where $i = Ni$ or $NiH$ obtained by fitting the equation to enthalpies mentioned in Fig. 8(a) and (b). The parameter values in the table can be directly used with a {100} oriented crystal. A tensor**

**transformation is required for other orientations (equation 5). Value of $C_{44}$ mentioned in Table 4 is used in such a case.**

| | | Ni | NiH |
|---|---|---|---|
| Heat capacity | $C_{P,0}$ (eV/(Ni atom.K)) | 0.0002722 | 0.0005251 |
| | $C_{P,T}$ (eV/(Ni atom.K$^2$)) | 0.000000047 | 0.00000032 |
| | $C_{P,\epsilon}$ (eV/(Ni atom.K)) | 0.000097 | 0.00017 |
| Elastic constants | $C_{11,0}$ $(\mathrm{eV.Å^{-3}})$ | 1.539 | 1.488 |
| | $C_{11,T}$ $(\mathrm{eV.Å^{-3}.K^{-1}})$ | -0.00094 | -0.00294 |
| | $C_{11,\epsilon}$ $(\mathrm{eV.Å^{-3}})$ | 0.00028 | 0.000939 |
| | $C_{11,T\epsilon}$ $(\mathrm{eV.Å^{-3}.K^{-1}})$ | -0.00233 | -0.000908 |
| | $C_{12,0}$ $(\mathrm{eV.Å^{-3}})$ | 0.9272 | 1.143 |
| | $C_{12,T}$ $(\mathrm{eV.Å^{-3}.K^{-1}})$ | 0.00026 | 0.00136 |
| | $C_{12,\epsilon}$ $(\mathrm{eV.Å^{-3}})$ | -0.0013 | -0.0045 |
| | $C_{12,T\epsilon}$ $(\mathrm{eV.Å^{-3}.K^{-1}})$ | 0.00294 | 0.0027 |

In Fig. 8(a), the red circles obtained at a particular temperature are found to be arranged in an arc or C-shape. High temperature causes the molar enthalpy to increase, and the curvature of the arc is lowered. The Pearson correlation coefficient $R^2$ obtain for the model is 0.98. A better result is obtained (see black plus symbols) by including the coupling parameters of equation (10) in the model. The $R^2$ for such a model is 0.99 for Ni phase. Similar conclusions are obtained for NiH where $R^2$ is 0.98. The $R^2$ value improves from 0.98 to 0.99 when the coupling parameters are included. Table 5 lists the hyper elastic model parameters.

When the Ni or the NiH crystal is rotated in the {111} direction, the same bulk properties estimated from {100} orientation along with the corrections can be used by applying the transformation rules. Fig. 8(c) and (d) that the enthalpy model with corrections included is consistent with the MD results.

### 4.4. Interfacial energy accounting for coupled chemical, strain and temperature effects

Now that the material properties (cohesive energy, lattice parameter, elastic constant, heat capacity, and thermal expansion coefficient) are available, we are in a position to determine the enthalpy and volume of interfacial systems.

For a periodic α/β interfacial system with no surface present, like the one shown in Fig. 2, equation (1) can be written as,

$$\underline{H}(T,P,x,\{l\},\lambda) = x_\alpha \underline{H_\alpha}(T,P,\{l\}_\alpha) + x_\beta \underline{H_\beta}\left(T,P,\{l\}_\beta\right) + A_{int}\lambda. \quad (11)$$

Here an additional term, so far undetermined, appears due to the interface. $\lambda$ is interfacial energy and $A_{int}$ is the interface area per Ni atom. Equation (11) also provides the lattice deformation in α and β phases, as discussed next. Two low-energy interfaces, namely, in the {100} and {111} directions are studied and the associated interface energy is computed.

The equilibrium dimension of the periodic system $\{l\}$ depends on the extent of hydride formation (see Fig. 2 and Fig. 3). Once again, assuming that $\underline{G} \approx \underline{H}$, minimization of $\underline{H}\left(T, l_{x,\alpha}, l_{y,\alpha}, l_{z,\alpha}, l_{z,\beta}\right)$ at constant temperature and pressure provides four coupled equations

$$\frac{d\underline{H}}{dl_{x,\alpha}} = 0, \quad (12)$$

$$\frac{d\underline{H}}{dl_{y,\alpha}} = 0,$$

$$\frac{d\underline{H}}{dl_{z,\alpha}} = 0,$$

$$\frac{d\underline{H}}{dl_{z,\beta}} = 0$$

which are solved simultaneously to determine the dimensions $\{l\}$ and the enthalpy of the relaxed system.

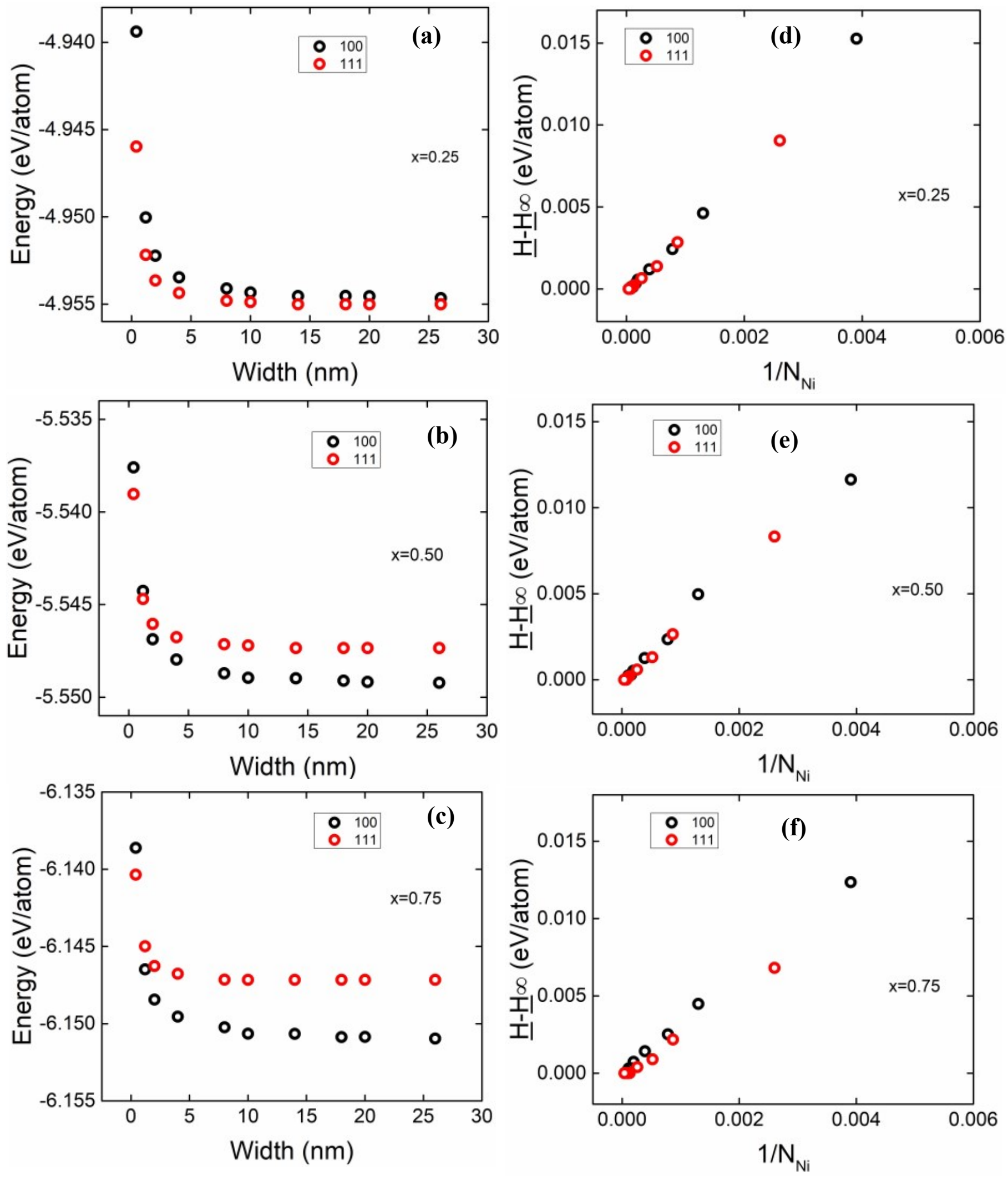

**Fig. 9.** Enthalpy as a function of the width of the Ni/NiH interfacial system. {100}, and {111}, and interfaces with (a) x=0.25, (b) x=0.50, and (c) x=0.75 at 300 K are considered. $\underline{H}$-$\underline{H}_{\infty}$ vs. $\frac{1}{N_{Ni}}$ in (d, e, f) is shown for {100}, and {111} planes for NiH bulk system at 300 K. Here, energy is reported in units of eV per Ni atom.

The estimation of $\lambda$ is addressed next. MD simulations in the NPT ensemble are utilized to determine the molar enthalpy and the dimension $\{l\}$. For the {100} interface, we have used supercells of different sizes, namely 4×4×2, 4×4×4, 4×4×40, 4×4×100, 4×4×180 and 4×4×260 unit cells to probe the size effect. In contrast, for the {111} direction, we employed 4×4×2, 4×4×4, 4×4×40, 4×4×100, 4×4×180 and 4×4×260 repeat units. The repeat unit information for the {111} interface is provided in the supplementary material in section (d) (Table S8). The molar enthalpy (eV/Ni atom) at 300 K is calculated for x=0.25, 0.50, and 0.75, as shown in Fig. 9(a, b, c). For a 4×4×180 supercell system, x=0.75 implies 25% of the unit cells (45 of the 180 unit cells in the z-direction) form the α-phase and the remaining the β-phase. Our MD simulations confirm that no dislocations are created and a coherent Ni-NiH lattice is present. Results for 400 and 500 K are presented in supplementary Fig. S5.

The horizontal axis of Fig. 9 represents the total equilibrium width of the periodic α/β interfacial system along z-direction, i.e., $l_{z,\alpha} + l_{z,\beta}$, as obtained from MD. The MD simulation results are shown in open circles for the {100} and {111} directions. The relative size of the interfacial to bulk region determines the interface contribution. Recall that in our notation $A_{int}$ is the interface area normalized by the number of Ni atoms in the system. Since $A_{int} \to 0$ when $l_{z,\alpha} + l_{z,\beta}$ is large, the molar enthalpy mainly contains contributions from the bulk regions. In such a case, equation (11) reduces to

$$\underline{H}(T, P, x, \{l\}, \lambda) = x_\alpha \underline{H}_\alpha(T, P, \{l\}_\alpha) + x_\beta \underline{H}_\beta\big(T, P, \{l\}_\beta\big). \qquad (13)$$

This is seen in in Fig. 9(a, b, c) where the molar enthalpy plateaus for 15 nm and larger widths. $\underline{H}$ becomes independent of the system size. The observed trends are: $\underline{H}_{111} < \underline{H}_{100}$ at x=0.25. Here the subscript refers to the interface. Note that the molar enthalpies lie within 1 meV/Ni atom of each other. The trends change beyond x=0.5, where $\underline{H}_{100} < \underline{H}_{111}$. At x=0.75, the molar enthalpies lie within 7 meV of each other. An important point emerges: these trends are attributed pure to only elastic effects, since the interface contribution is negligible. For smaller supercell width, a crossover from {111} interface to a {100} interface when we view the molar enthalpy in Fig. 9. Thus, size and chemo-thermo-mechanical effects are coupled.

Denoting $\underline{H}_{\infty}$ as the largest supercell energy (26 nm width, in Fig. 9(a, b, c). The excess enthalpy $\underline{H}$-$\underline{H}_{\infty}$ is used to determine the interface term for the smaller supercells. The interface energy must be positive, as the molar enthalpy $\underline{H}$ tends to increase for small supercell widths, as seen in Fig. 9. Fig. 9(d, e, f) reveal that $\underline{H}$ vs. $A_{int}$ is linear and the slope determines $\lambda$,

$$\lambda = \frac{1}{A_{int}}\left(\underline{H} - \underline{H}_{\infty}\right). \tag{14}$$

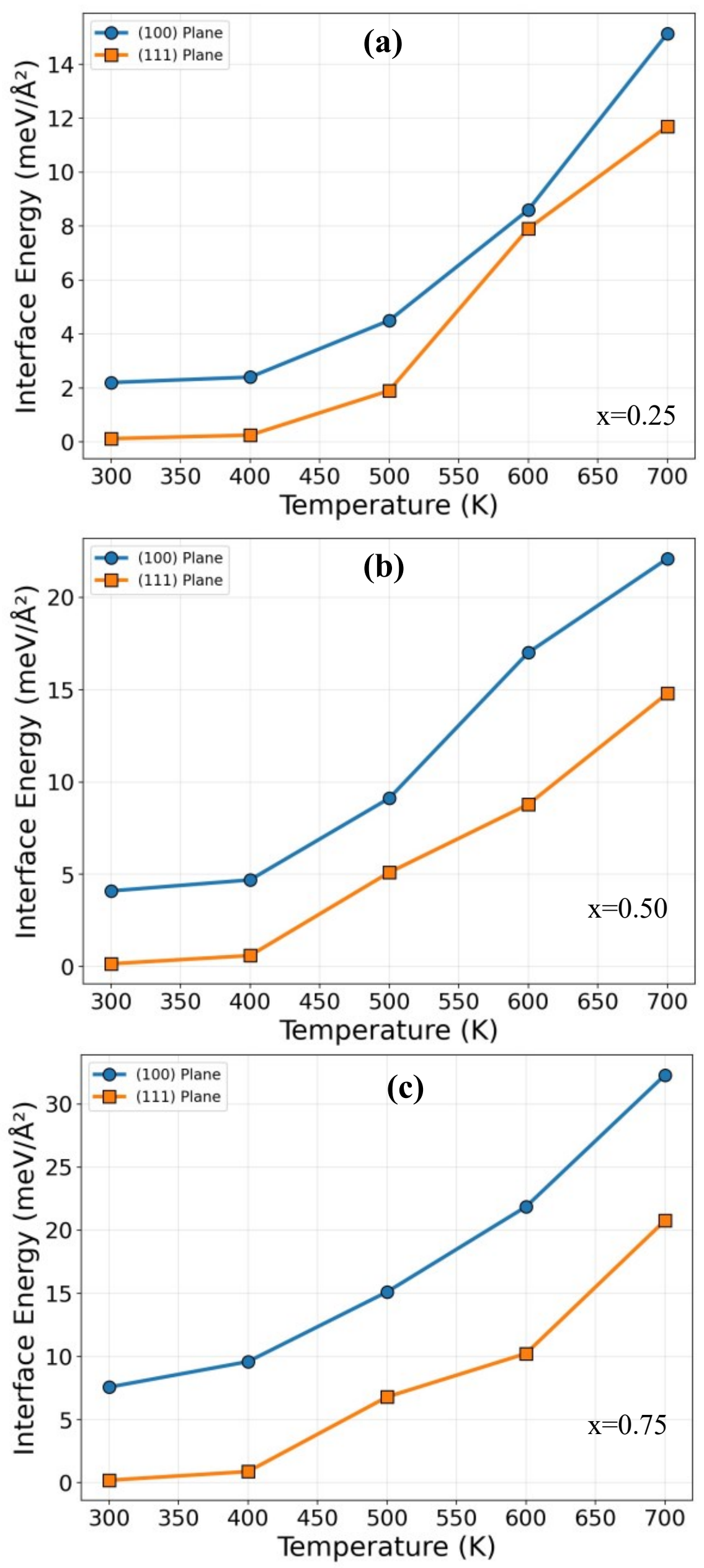

**Fig. 10.** Interfacial energy vs. temperature for {100} and {111} interfaces. Here x is (a) 0.25, (b) 0.50, and (c) 0.75.

Effect of temperature on the interface energy is shown in Fig. 10. The interface energy increases superlinearly with respect to temperature. The {111} interface continues to have lower energy than the {100} interface. Based on these discussions, it becomes clear that interfacial energy can play important role in thin films and small nanoparticle systems. For instance, the preferred hydrogen growth direction may be dictated by the interfacial energy.

### 4.5. Size dependent enthalpy and the lattice parameters for the hydride system

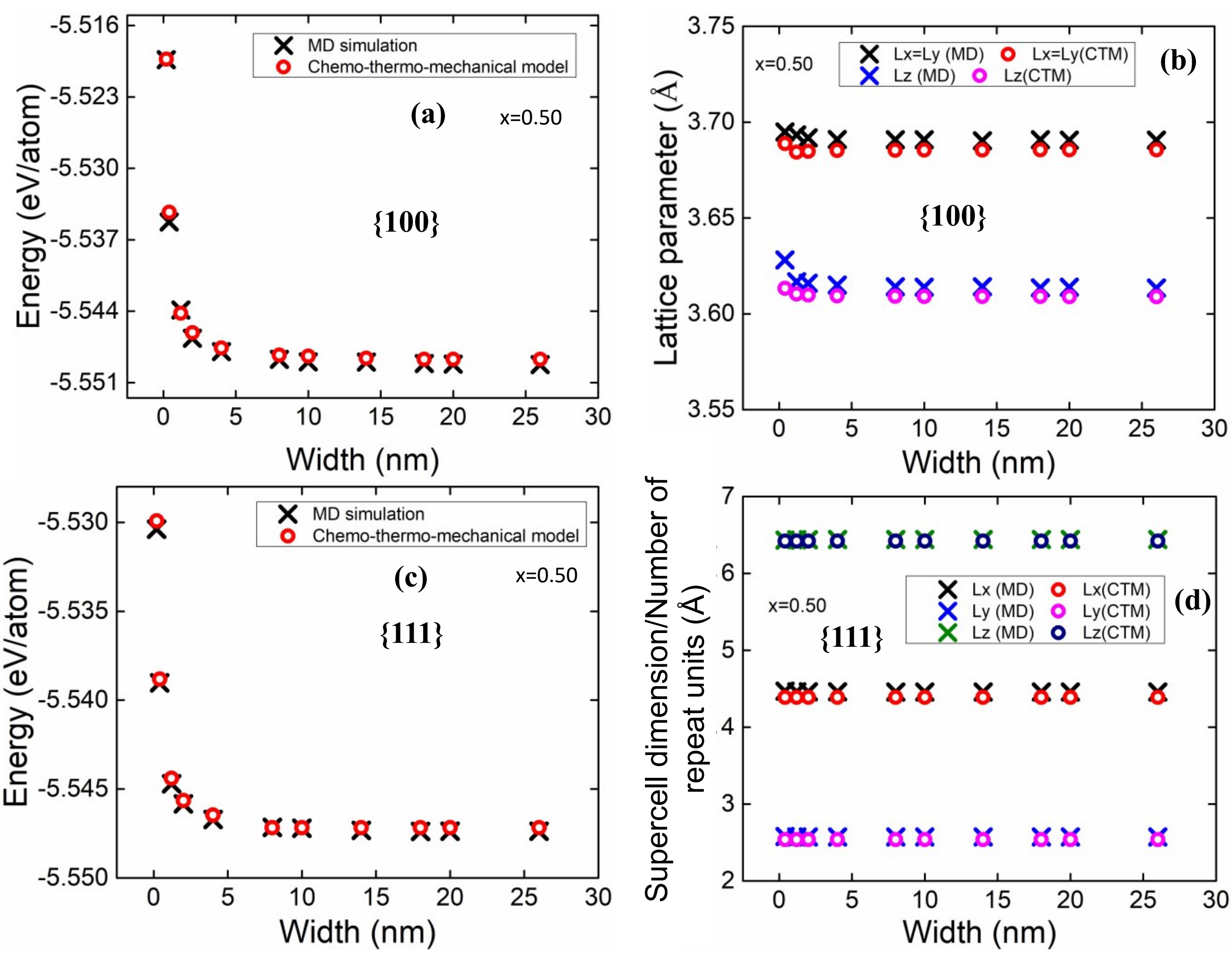

**Fig. 11.** Comparison of MD simulation results with the chemo-thermo-mechanical model at 300 K for hydrogen concentration $x = 0.50$: (a) energy and (b) deformation along the z-direction for the {100} oriented crystal; (c) energy and (d) deformation along the z-direction for the {111} oriented crystal. Here, CTM= Chemo-thermo-mechanical. Here, energy is reported in units of eV per Ni atom.

Hydride formation is associated with heat release and volume change. A comparison between the total molar enthalpy and equilibrium lattice parameter from MD (in the NPT ensemble) and the chemo-thermo-mechanical model (using equation (12)) is made in Fig. 11. The MD and model results are shown in cross and open circle symbols, respectively. Overall excellent agreement is obtained highlighting the model ability to capture size effect appropriately.

In order to form a coherent lattice in a {100}-oriented crystal, the α phase undergoes expansion in the x- and y-direction, whereas the β phase undergoes contraction. This directly impacts the dimensions of the respective phases in the z-direction as well, viz., the α phase contracts while the β phase expands in the z-direction. The effective lattice parameter, calculated by box dimension normalized by the number of unit cells, is plotted in Fig. 11(b).

The same exercise is performed for the {111} oriented crystal. In Fig. 11(d), we plot the supercell dimension divided by the number of repeat units (see details of the repeat unit in Table S8 in Supplementary Material). As we can see in Table S8, the repeat unit dimension is 4.33×2.5×6.12 Å for fully relaxed Ni(111), and 4.61×2.66×6.52 Å for fully relaxed NiH(111). In Figure 11(d), we find that the dimension per repeat unit is nearly 4.47×2.57×6.44 Å for the relaxed Ni/NiH(111) interface. Overall, Fig. 11(c) and (d) confirm that the chemo-thermo-mechanical model is able to correctly predict the enthalpy and volume changes.

### 4.6. Surface energy

Comparing the lattice parameters for {100} and {111} oriented crystals (Figs. 11(b)-(d)), we conclude that these are largely dictated by the elastic energy terms. We anticipate that surface energy will behave in a manner similar to the interfacial energy, i.e., it is less significance than the elastic terms.

First, let us consider the surface energy at 300 K, which is calculated separately for Ni and NiH slabs with {100} and {111} surfaces. A supercell size 4×4×4 is used. The lattice parameters were allowed to change according to temperature using molecular dynamics simulations an NPT ensemble. For NiH, the surface is hydrogen terminated, i.e., adsorbed H is present. The surface energy is calculated as

$$\gamma = \frac{\underline{H}_{slab} - \underline{H}_{bulk}}{A_{surface}} \quad (15)$$

where $\underline{H}_{slab}$ is energy of the slab, $\underline{H}_{bulk}$ is the energy of the bulk, and $A_{surf}$ $(\text{Å}^2/Ni\ atom)$ is the surface area of the slab. Based on these calculations, $\gamma_{100,\alpha}$ and $\gamma_{111,\alpha}$ are 0.1272 and 0.118 (eV/$\text{Å}^2$), respectively, whereas $\gamma_{100,\beta}$ and $\gamma_{111,\beta}$ are 0.0926 and 0.08696 (eV/$\text{Å}^2$), respectively. In literature, the surface energy of Ni {100} and {111} planes are reported to lie between 0.137 to 0.152 (eV/$\text{Å}^2$) and 0.1211 to 0.1255 (eV/$\text{Å}^2$), respectively [61][62]. The results show that $\gamma_{111} < \gamma_{100}$. Fig. 12(a) shows the surface energy as a function of temperature. The {111} surface has lower energy than {100} plane, and trends reverses for NiH as shown in Fig. 12(b). Fig. 12 confirms that the surface energy term is indeed small.

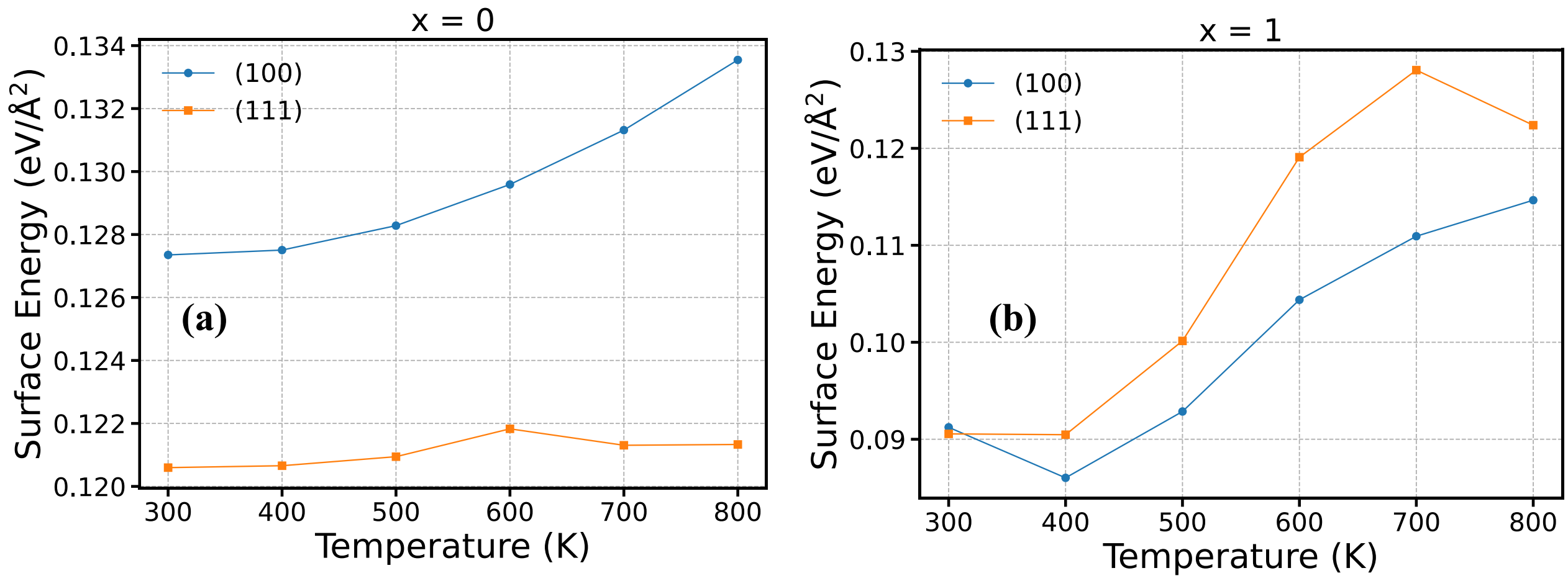

**Fig. 12**: Surface energy as function of temperature (300 to 800 K) for (a) Ni and (b) NiH system.

## 5. Conclusion

In solid-state hydrogen storage materials, volume changes and heat release are common during the absorption/desorption process. A thermodynamic model that accounts for the coupled chemical, thermal and mechanical effects is required. In addition, applications often use nanometer to micron sized particles, hence, size effect also needs to be considered. A size-dependent chemo-thermo-mechanical model is systematically developed for nickel hydride.

The main findings are that when a metal/metal hydride interface is formed the material properties are directly influenced by the chemo-thermo-mechanical effects. Corrections to the material properties is required as a function of the extent of hydride formation, temperature and elastic strain. The interfacial and surface energy terms are small, and become relevant only when the dimensions are less than 20 nm. Surprisingly we observe that for widths in excess of 20 nm, {111} interface is lower in energy than the {100} interface when x<0.25 (in $NiH_x$), and this trend reverses for higher x values. Such behavior can be explained in terms of the anisotropic elastic constants. Consequently, hydride formation should initially proceed in the {111} direction, and later switch to a {100} direction.

Although, we are unaware of any experimental reports involving anisotropic hydride growth in Ni, in Ref. [15], we did observe anisotropic growth behavior in Ni nanoparticles using Monte Carlo simulations. It was observed that in Ni nanoparticles the hydride growth initially happens at the {111} interface, and at higher hydrogen concentrations, growth proceeds at the {100} interface. The present study provides an explanation, that these effects arise primarily due to the anisotropic elastic constants, causing the {111} interface to be energetically more favorable, as long as $x < 0.25$. At higher hydrogen concentration, the {100} interface becomes more favorable.

It is well known that in several metal hydride systems, the hydride growth can be anisotropic. Li et al. utilized transmission electron microscopy (TEM) to demonstrate that hydride growth in palladium (Pd) nanocrystals occurs more significantly at the {111} plane compared to the {100} plane [63]. An experimental study on Pd/Zr nanoparticles demonstrates {111}-preferred growth at low loading, influenced by strain, as observed by TEM analysis [64,65]. Narayan et al. [38] shows β phase growth on Pd nanoparticle by using scanning transmission electron microscopy (STEM) at first {111} plane than {100) plane.

Our approach paves the way for understanding the thermodynamics of hydride formation. To account for these aspects in the Ni system, in general one can develop a chemo-thermo-mechanical enthalpy model, denoted as $\underline{H}(x, T, P, l, \lambda)$, that captures the combined influence of extent of hydride formation $x$, temperature $T$, pressure $P$, length dimensions ($l$), the α/β interface energy $\lambda$, and the α and β surface energy terms. Here the α/β interface refers to the metal/hydride interface.

## 6. Acknowledgements

AC would like to acknowledge the Computer Centre at the Indian Institute of Technology Bombay for providing computational support. This work was also supported by the National Supercomputing Mission, India, through the DST/NSM/R&D_HPC_Applications grant (2021/02). AC acknowledges funding support from the Centre of Excellence in Oil, Gas and Energy (CoEOGE) at the Indian Institute of Technology Bombay. The authors sincerely thank Mr. Annaji Rajiv Kumar Tompala from Hindustan petroleum Corporation Limited (HPCL) for helpful discussions. The authors acknowledge the help provided by Dr. Poonam Parkar and Mr. Divy Vankar in the DFT calculations provided in the supplementary material.

## 7. Data Availability Statement

The data that support the findings of this study are available in the supplementary material of this article.

## 8. Author contributions

SS and AC conceptualized the work; SS conducted the investigation; SS and AC jointly wrote the manuscript.